\documentclass[a4paper,onecolumn,11pt,unpublished]{quantumarticle}
\pdfoutput=1
\usepackage[utf8]{inputenc}
\usepackage[english]{babel}
\usepackage[T1]{fontenc}
\usepackage{amsmath}
\usepackage{amssymb}
\usepackage{amsfonts}
\usepackage[colorlinks,linkcolor=blue,urlcolor=blue,citecolor=blue]{hyperref}
\usepackage{bm}

\usepackage[sort&compress,numbers]{natbib}

\usepackage{physics}
\usepackage{mathtools}

\title{Low Depth Color Code Circuits with CXSWAP gate}
\author{Satoshi Yoshida}
\orcid{0000-0002-0521-5209}
\email{satoshiyoshida.phys@gmail.com}
\affiliation{Department of Physics, Graduate School of Science, The University of Tokyo, 7-3-1 Hongo, Bunkyo-ku, Tokyo 113-0033, Japan}
\affiliation{Google Quantum AI, Tokyo, Japan}
\author{Craig Gidney}
\affiliation{Google Quantum AI, Santa Barbara, California 93117, USA}
\author{Matt McEwen}
\affiliation{Google Quantum AI, Santa Barbara, California 93117, USA}
\author{Adam Zalcman}
\orcid{0000-0002-2585-2424}
\affiliation{Google Quantum AI, Santa Barbara, California 93117, USA}
\affiliation{Google Quantum AI, Tokyo, Japan}
\begin{document}

\begin{abstract}
    We present two new types of syndrome extraction circuits for the color code. Our first construction, which after Ref.~\cite{McEwen2023relaxinghardware} we call the \textit{semi-wiggling} color code, promises to mitigate leakage errors by periodically interchanging the roles of bulk data and measurement qubits. The second construction reduces circuit depth relative to Ref.~\cite{gidney2023new} by employing the CXSWAP gate instead of CNOT. This optimization leads to $\sim 10\%$ improvement in teraquop footprint under the uniform error model with the physical error rate $p=0.1\%$.
\end{abstract}

\maketitle

\section{Introduction}

Recent developments in quantum technology enable experiments to demonstrate exponential error suppression using quantum error correction~\cite{google2023suppressing, google2025quantum}.
These experiments use the surface code which exhibits a high threshold and requires only 2-dimensional nearest-neighbor (2DNN) connectivity. However, it also incurs a large qubit overhead, especially for fault-tolerant implementations of logical operations~\cite{campbell2017roads}.

The color code is a topological quantum error correcting (QEC) code that, like the surface code, can be implemented in the 2DNN architecture~\cite{bombin2006topological}.
It supports transversal implementation of all Clifford gates and requires fewer qubits than the surface code to achieve the same distance.
It can also be used with the surface code to reduce the cost of producing magic states for fault-tolerant realizations of non-Clifford operations~\cite{itogawa2024even, gidney2024magic}.

However, syndrome extraction is more difficult in the color code due to the higher weight of its stabilizer generators.
While the bulk stabilizer generators in the surface code have weight four, the bulk stabilizers in the color code on the honeycomb lattice have weight six, which leads to longer syndrome extraction circuit and a lower threshold.
Recently, the error correction circuits for the color code have been improved~\cite{gidney2023new, takada2024improving} and experimentally realized~\cite{lacroix2024scaling}.
Nevertheless, there is still room for improvement compared to the surface code circuits which have attracted greater attention~\cite{McEwen2023relaxinghardware, fowler2012surfacecode, paetznick2013quantum, litinski2019game, chao2020optimization, lee2022lattice, srivastava2023encoder, wang2011surface, debroy2024lucisurfacecodedropouts, leroux2024snakesladdersadaptingsurface}.

Here, we develop new quantum circuits for the color code by employing techniques of Ref.~\cite{McEwen2023relaxinghardware} for the surface code circuits (which is experimentally implemented in Ref.~\cite{eickbusch2024demonstrating}) to the circuits proposed in Ref.~\cite{gidney2023new}.
We show that the depth of the error correction circuit can be reduced by replacing the CNOT gates with the CXSWAP gates, which is defined as the product of the CNOT gate and the SWAP gate, and is easy to implement in some superconducting qubit architectures~\cite{foxen2020demonstrating}.
We numerically demonstrate that this construction offers $\sim 10\%$ reduction in the number of qubits necessary to achieve the logical error rate $10^{-12}$ under the uniform error model with the physical error rate $p=0.1\%$ than the corresponding circuits using CNOT gates, assuming that the CXSWAP gate and the CNOT gate have the same error rates.
We also propose an error correction circuit which periodically interchanges the roles of data and measurement qubits in the bulk which enables leakage mitigation schemes that rely on the reset gate.
Similarly to Ref.~\cite{McEwen2023relaxinghardware}, we achieve the leakage mitigation without increasing the depth of the circuit, contrary to several previous approaches requiring adding minimal operations~\cite{fowler2013coping,ghosh2015leakage,suchara2015leakage,brown2019leakage,battistel2021hardware,mcewen2021removing,miao2023overcoming}.

The rest of this paper is organized as follows.
Section~\ref{sec:circuit_constructions} shows the construction of the error correction circuits for the color code and Section~\ref{sec:benchmarking} analyzes their performance.
Section~\ref{sec:conclusion} concludes this work.
In addition, Appendix~\ref{appendix_sec:error_model} shows the definition of the error model used in our numerical simulations and Appendix~\ref{appendix_sec:numerical_result} shows detailed numerical results.

\section{Circuit constructions}
\label{sec:circuit_constructions}

Ref.~\cite{McEwen2023relaxinghardware} introduces techniques for directly designing time-dynamic QEC \textit{circuits} as an effective alternative to the traditional approach of going through the intermediate step of designing static QEC \textit{codes}. The techniques focus on detectors and detecting regions instead of the stabilizer group and its generators. In addition, Ref.~\cite{McEwen2023relaxinghardware} employs the new approach to develop three types of syndrome extraction circuits for the surface code:
\begin{itemize}
    \item Circuits that can be embedded in the hexagonal grid instead of the square grid.
    \item Circuits that periodically flip the roles of data and measurement qubits.
    \item Circuits that use ISWAP gates instead of CNOT or CZ gates.
\end{itemize}
The first circuit type reduces connectivity requirements of the surface code which makes it possible to run it on platforms with connectivity constraints more stringent than those of the square grid as well as on some devices with broken components. The second type of circuit enables leakage mitigation schemes realized by specialized reset gates to be applied to all qubits. The third type allows the surface code to be executed using the ISWAP gate rather than the more conventional CNOT or CZ gates. In some architectures, such as those based on superconducting qubits, the ISWAP gate is easier to calibrate with high fidelity~\cite{foxen2020demonstrating}.

We apply these techniques to the midout and superdense color code circuits presented in Ref.~\cite{gidney2023new}.
The midout color code circuit is already embedded in a hexagonal grid, so we focus on color code analogues for the second and third type of circuit. Our first construction periodically flips the roles of data and measurement qubits, like the wiggling circuits in Ref.~\cite{McEwen2023relaxinghardware}, albeit only in the bulk, so we refer to these circuits as the \textit{semi-wiggling} circuits.
It is based on the midout color code circuits.
Our second construction uses the CXSWAP gate which is equivalent to ISWAP under single-qubit Clifford operations
\begin{align}
    \mathrm{CXSWAP} = (S^\dagger \otimes HS^\dagger) \, \mathrm{ISWAP} \, (H\otimes I),
\end{align}
but is more convenient to analyze, because, like CNOT and unlike ISWAP, it sends $X$- and $Z$-type Pauli operators to Pauli operators of the same type under conjugation. Our second construction comes in two variants: one based on the midout color code circuits and one based on the superdense color circuits.
We discover that it is possible to reduce the depth of the CXSWAP circuits relative to the conventional CNOT circuits.

\begin{figure}
    \centering
    \includegraphics[width=\linewidth]{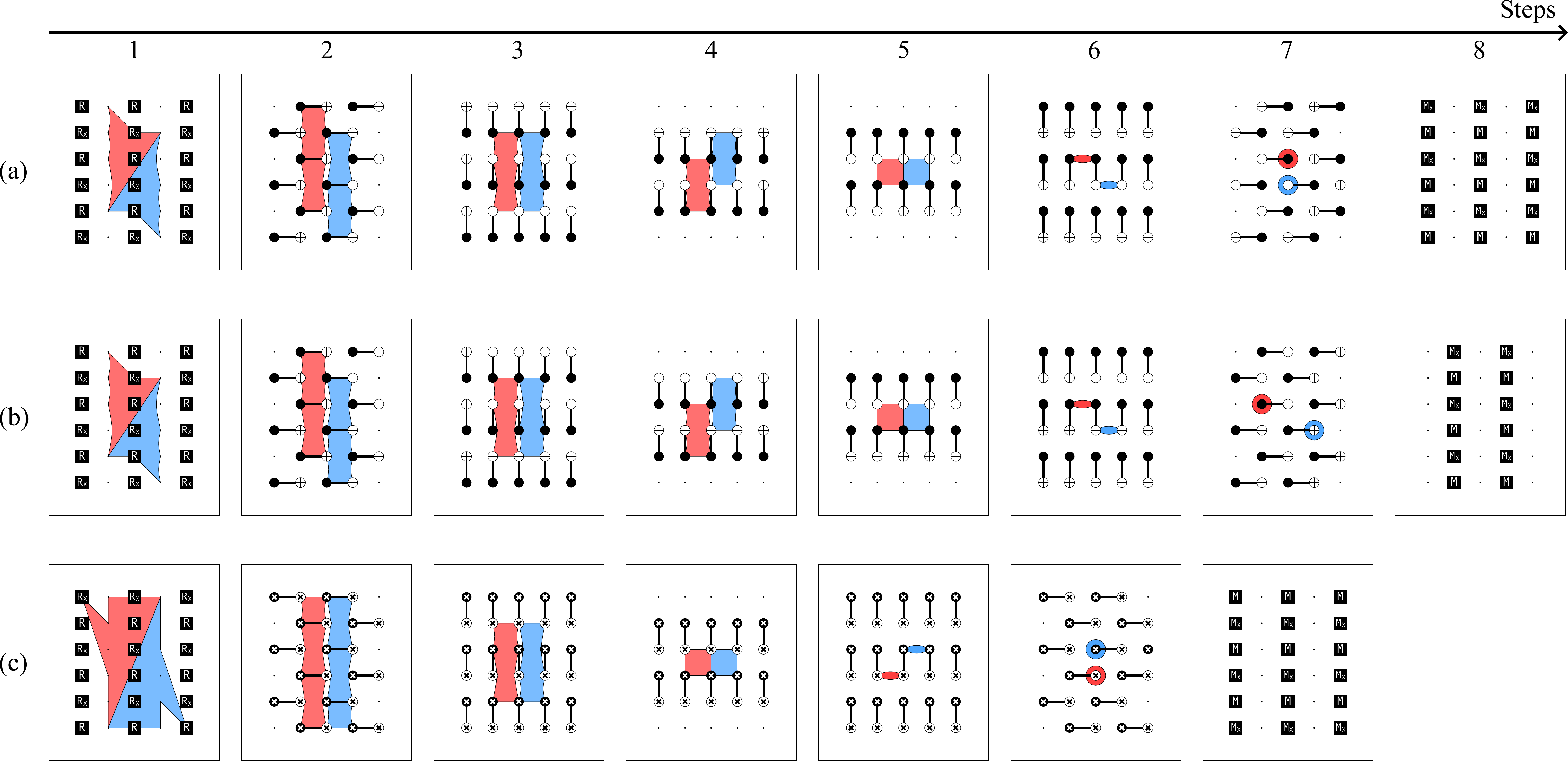}
    \caption{(a) One cycle of a bulk fragment of the midout circuit consisting of a reset step, 6 CNOT steps, and a measurement step.
    The X basis detector (red) and Z basis detector (blue) start from a 6-body stabilizer in the reset step, shrink to a one-body stabilizer in the CNOT layers, and are measured in the measurement step.
    The stabilizer for the color code appears after the 4th step.\\
    (b) One cycle of a bulk fragment of the semi-wiggling midout circuit.
    CNOT direction in the 7th step is flipped relative to the original midout circuit causing a one-column shift of measurement qubits.\\
    (c) One cycle of a bulk fragment of the CXSWAP midout circuit.
    This circuit is obtained by contracting the middle two CNOT steps of (b) to one CXSWAP step in the 4th step, and replacing the other CNOT steps with the CXSWAP steps followed by propagating the SWAP gates to the reset and measurement steps.}
    \label{fig:bulk_midout_circuit}
\end{figure}

\subsection{Semi-wiggling midout circuit}

\begin{figure}
    \centering
    \includegraphics[width=.7\linewidth]{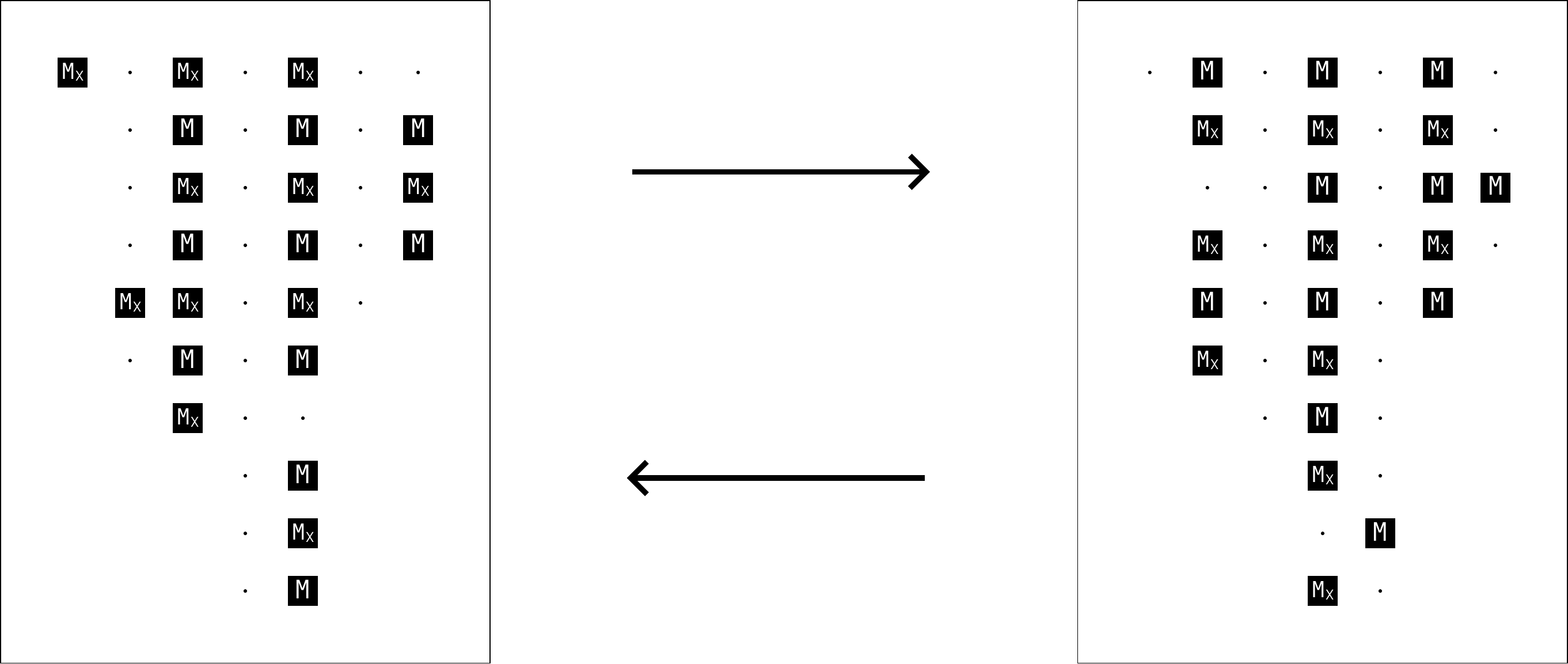}
    \caption{The locations shown in the diagrams on the left and right are alternately used as measurement qubits in each cycle of the semi-wiggling midout circuit.}
    \label{fig:boundary_wiggling_measurement}
\end{figure}

We construct our semi-wiggling midout circuit based on the midout circuit shown in Ref.~\cite{gidney2023new}.
The original midout circuit is shown in Fig.~\ref{fig:bulk_midout_circuit}~(a), where the color code stabilizers appear in the middle of the error correction circuit.
In the bulk of the semi-wiggling circuit, measure and data qubits are flipped in each measurement step as shown in Fig.~\ref{fig:boundary_wiggling_measurement}, which can be used to mitigate leakage errors.

The key idea behind this construction can be described as follows.
In the original midout circuit, the final CNOT step shrinks a 2-body stabilizer to a 1-body stabilizer.
Therefore, by flipping the CNOT direction of this step, we can shrink 2-body stabilizer to the next column, which becomes the measurement qubit, see Fig.~\ref{fig:bulk_midout_circuit}~(b).
However, this construction does not work perfectly around the boundary, which causes an imperfect flipping of the measurement and data qubits, see Fig.~\ref{fig:boundary_wiggling_measurement}.
Also, this construction does not shift the stabilizer in the final measurement layer, while the walking surface code circuit~\cite{McEwen2023relaxinghardware} shifts the stabilizer in the final measurement layer, which moves the stabilizer in each error correction cycle.
This is why we call this circuit a \emph{semi}-wiggling midout circuit.

\subsection{CXSWAP midout circuit}

We next describe a CXSWAP circuit based on the midout circuit from Ref.~\cite{gidney2023new}.
The original midout circuit is shown by Fig.~\ref{fig:bulk_midout_circuit}~(a).
Our main strategy is to use the following circuit identities:
\begin{align}
\label{eq:cxswap_two_cnot}
    \mathrm{CXSWAP}_{1,2} &= \mathrm{CNOT}_{1,2}\mathrm{CNOT}_{2,1},\\
\label{eq:cxswap_propagate_1}
    \mathrm{CNOT}_{1,2} &= \mathrm{CXSWAP}_{2,1} \mathrm{SWAP}_{1,2},\\
\label{eq:cxswap_propagate_2}
    \mathrm{CNOT}_{1,2}&= \mathrm{SWAP}_{1,2} \mathrm{CXSWAP}_{1,2} .
\end{align}
We first apply Eq.~\eqref{eq:cxswap_two_cnot} to the middle two steps [the 4th and 5th steps in Fig.~\ref{fig:bulk_midout_circuit}~(a)] to replace them with one CXSWAP step [the 4th step in Fig.~\ref{fig:bulk_midout_circuit}~(c)].
Then, we apply Eq.~\eqref{eq:cxswap_propagate_1} to the first two CNOT steps [the 2nd and 3rd steps in Fig.~\ref{fig:bulk_midout_circuit}~(a)] to replace them with CXSWAP and SWAP gates, and propagate SWAP gates to the reset step.
Similarly, we apply Eq.~\eqref{eq:cxswap_propagate_2} to the last two CNOT steps [the 6th and 7th steps in Fig.~\ref{fig:bulk_midout_circuit}~(a)] to replace them with CXSWAP and SWAP gates, and propagate SWAP gates to the measurement step.
The depth of the circuit is reduced from 8 to 7, which contributes to a lower logical error rate as shown in Section~\ref{sec:benchmarking}. None of the steps produces the midout state.

\begin{figure}
    \centering
    \includegraphics[width=\linewidth]{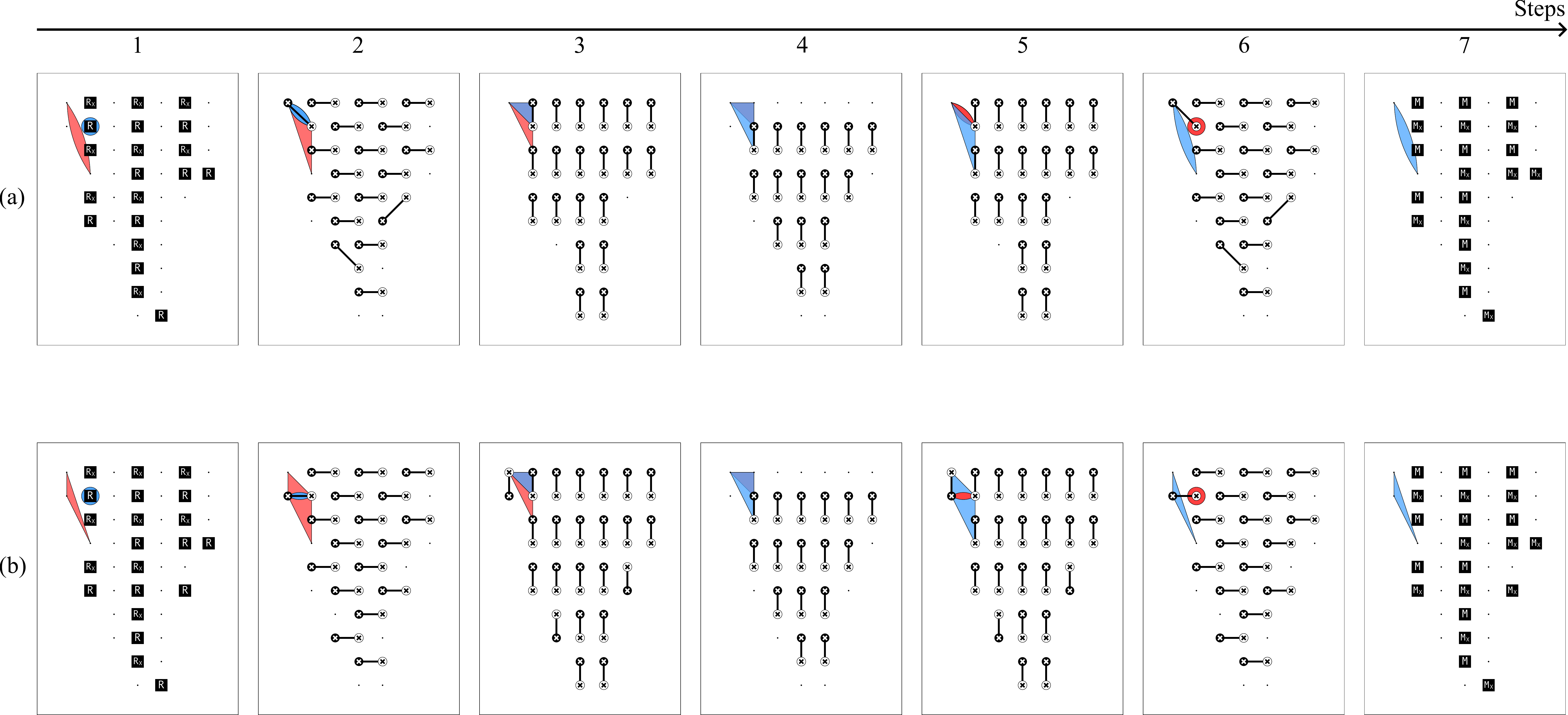}
    \caption{(a) The construction of the CXSWAP midout circuit shown in Fig.~\ref{fig:bulk_midout_circuit}~(c) generates diagonal gates on the boundary in the 2nd and 6th steps.\\
    (b) The diagonal gates in the 2nd step of (a) are replaced with the horizontal and vertical gates in the 2nd and 3rd steps, respectively.
    Similarly, the diagonal gates in the 6th steps of (a) are replaced with the vertical and horizontal gates in the 5th and 6th steps, respectively.}
    \label{fig:boundary_cxswap_midout_circuit}
\end{figure}

The above procedure gives rise to CXSWAP gates applied diagonally on the boundary, which break connectivity constraints of the 2-dimensional nearest-neighbor (2DNN) architecture [see Fig.~\ref{fig:boundary_cxswap_midout_circuit}~(a)].
To address this problem, we add auxiliary qubits on the boundary and decompose the diagonal CXSWAP gate into horizontal and vertical CXSWAP gates [see Fig.~\ref{fig:boundary_cxswap_midout_circuit}~(b)].
This requires $O(d)$ auxiliary qubits, which is an asymptotically vanishing fraction of the total number of qubits $O(d^2)$ as the code distance $d$ grows.

\subsection{CXSWAP superdense circuit}

\begin{figure}
    \centering
    \includegraphics[width=.7\linewidth]{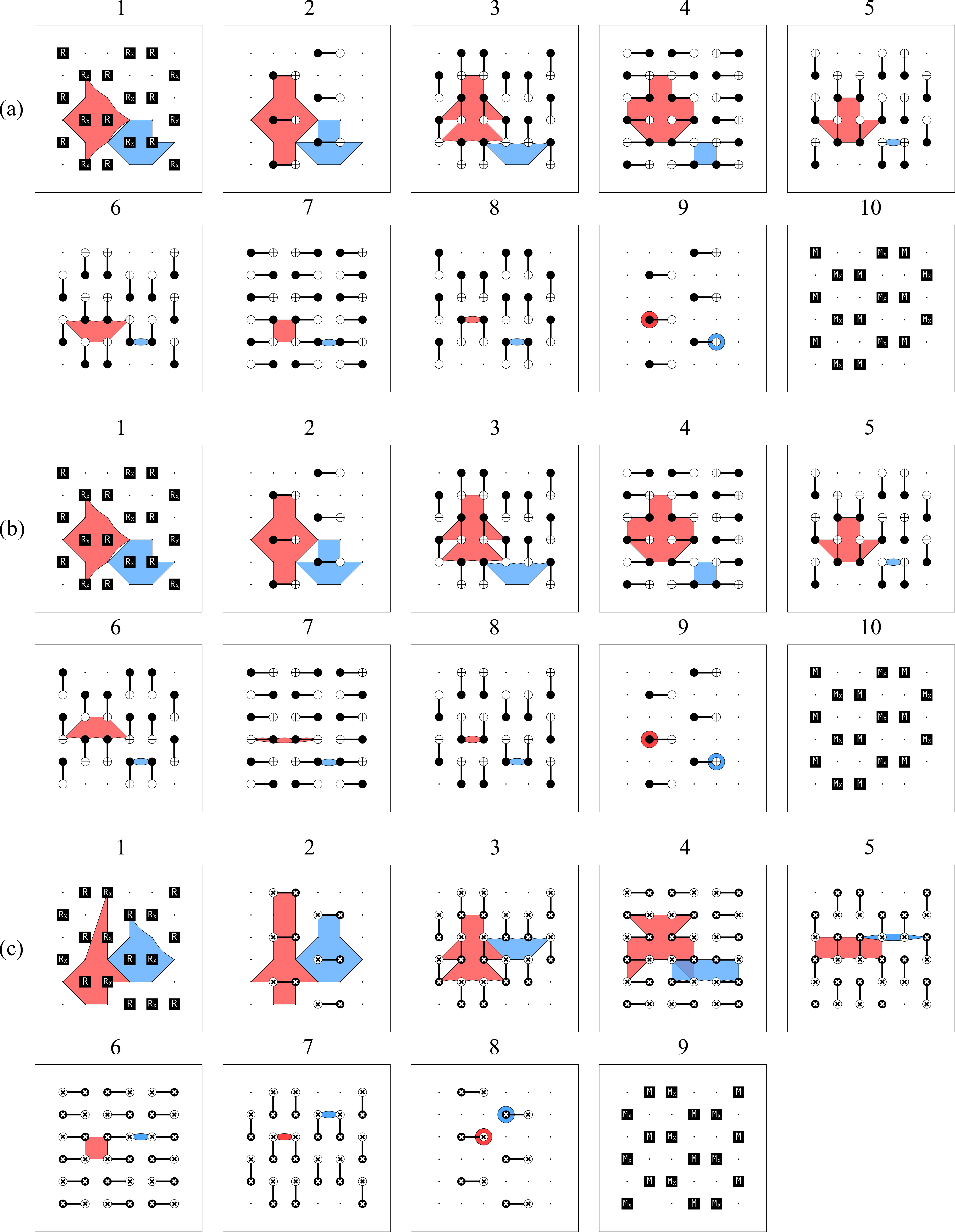}
    \caption{(a) One cycle of a bulk fragment of the original superdense circuit consisting of a reset step, 8 CNOT steps, and a measurement step.
    The Z detector shrinks to a 2-body stabilizer in the first 4 CNOT steps, and the X detector shrinks to a 2-body stabilizer in the following 3 CNOT steps.
    The last CNOT step shrinks the 2-body stabilizer to a 1-body stabilizer, which is measured in the last step.\\
    (b) A modified superdense circuit that is obtained by swapping the 6th and 8th steps in (a).\\
    (c) One cycle of a bulk fragment of the CXSWAP superdense circuit.
    This circuit is obtained by contracting the middle two CNOT steps of (b) to one CXSWAP step in the 5th step, and replacing the other CNOT steps with the CXSWAP steps followed by propagating the SWAP gates to the reset and measurement steps.}
    \label{fig:bulk_superdense_circuit}
\end{figure}

We also construct a CXSWAP circuit using the superdense circuit presented in Ref.~\cite{gidney2023new}.
A bulk fragment of the superdense circuit is shown in Fig.~\ref{fig:bulk_superdense_circuit}~(a).
The steps 3-5 shrink the Z detector to a 2-body stabilizer, the steps 6-8 shrink the X detector to a 2-body stabilizer, and the order of CNOT gates in the steps 3-5 (6-8) can be chosen arbitrarily.
We swap the 6th and 8th steps to apply a similar strategy as used in the CXSWAP midout circuit [see Fig.~\ref{fig:bulk_superdense_circuit}~(b)].
Similarly to the CXSWAP midout circuit, we construct a CXSWAP superdense circuit in the bulk [see Fig.~\ref{fig:bulk_superdense_circuit}~(b)].
The depth of the circuit is reduced from 10 to 9, which contributes to a lower logical error rate as shown in Section~\ref{sec:benchmarking}.

\begin{figure}
    \centering
    \includegraphics[width=\linewidth]{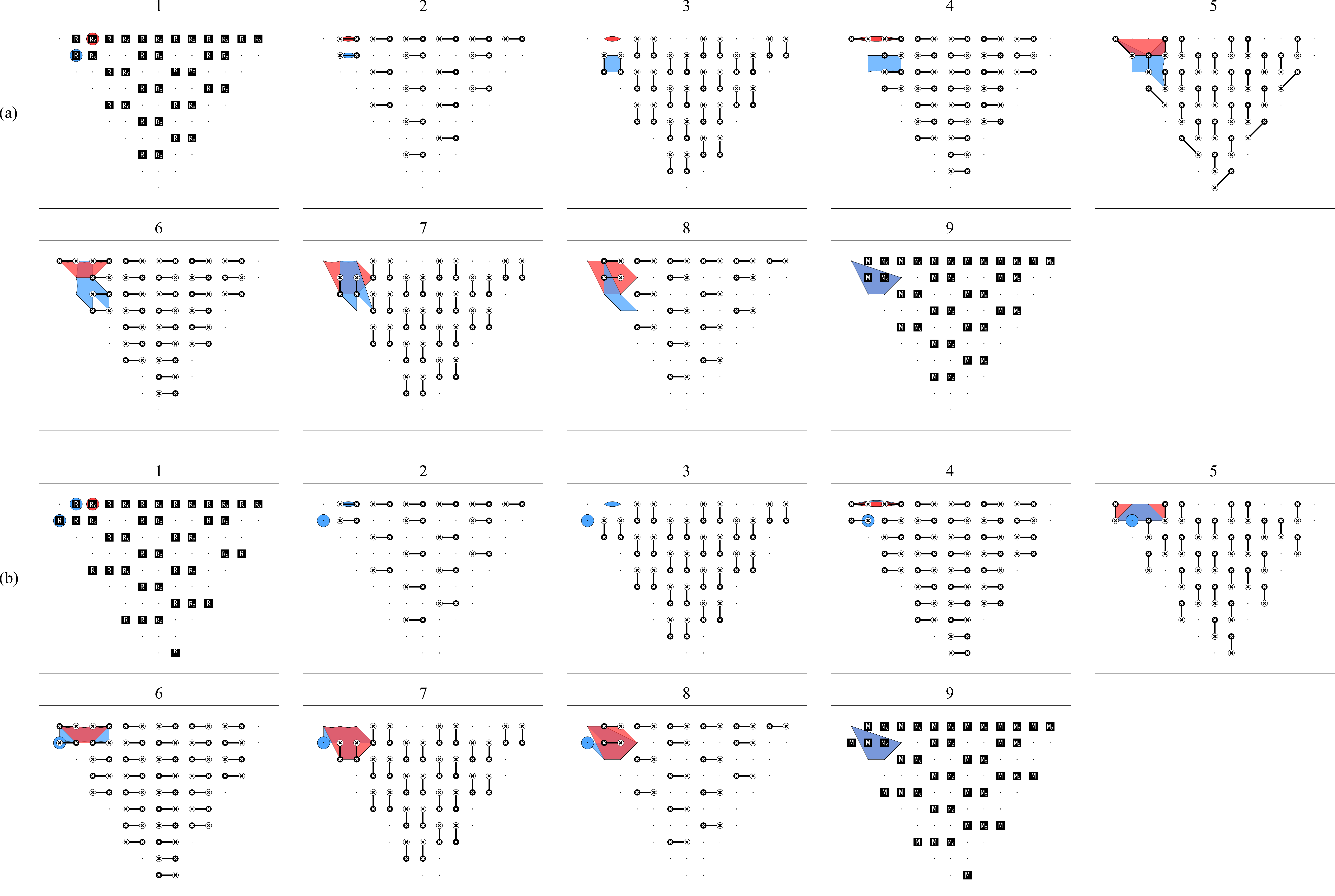}
    \caption{(a) The construction of the CXSWAP midout circuit shown in Fig.~\ref{fig:bulk_superdense_circuit}~(c) generates diagonal gates on the boundary in the 5th step.\\
    (b) The diagonal gates in the 5th step of (a) are replaced with the horizontal and vertical gates in the 4th, 5th, and 6th steps.}
    \label{fig:boundary_cxswap_superdense_circuit}
\end{figure}

The direct application of Eqs.~\eqref{eq:cxswap_two_cnot}--\eqref{eq:cxswap_propagate_2} give rise to diagonal gates on the boundary similarly to the CXSWAP midout circuit [see Fig.~\ref{fig:boundary_cxswap_superdense_circuit}~(a)].
We remove them by introducing auxiliary qubits on the boundary and decomposing the diagonal CXSWAP gates into horizontal and vertical CXSWAP gates [see Fig.~\ref{fig:boundary_cxswap_superdense_circuit}~(b)].

\section{Benchmarking}
\label{sec:benchmarking}

We numerically estimate the logical error rate of our circuits as follows.
We simulate a circuit to prepare a logical $\ket{0}$ ($\ket{+}$) state, run the error correction circuits for $4d$ rounds, and measure in the logical $Z$ ($X$) basis, where $d$ is the distance of the code.
We count the number of failed shots, repeat the simulation until the number of failed shots reaches $1000$, and estimate the failure probability by dividing the number of failed shots by the total number of shots.
We evaluate the logical error rate per round by dividing the failure probability by $4d$.
We conduct the numerical simulation under the uniform error model described in Appendix~\ref{appendix_sec:error_model}.
We use the stabilizer circuit simualtor {\tt Stim}~\cite{gidney2021stim} and the decoder {\tt chromobius}~\cite{gidney2023new, chromobius} based on the M\"{o}bius decoder~\cite{sahay2022decoder}.

\begin{figure}
    \centering
    \begin{minipage}{0.49\linewidth}
        \includegraphics[width=\linewidth]{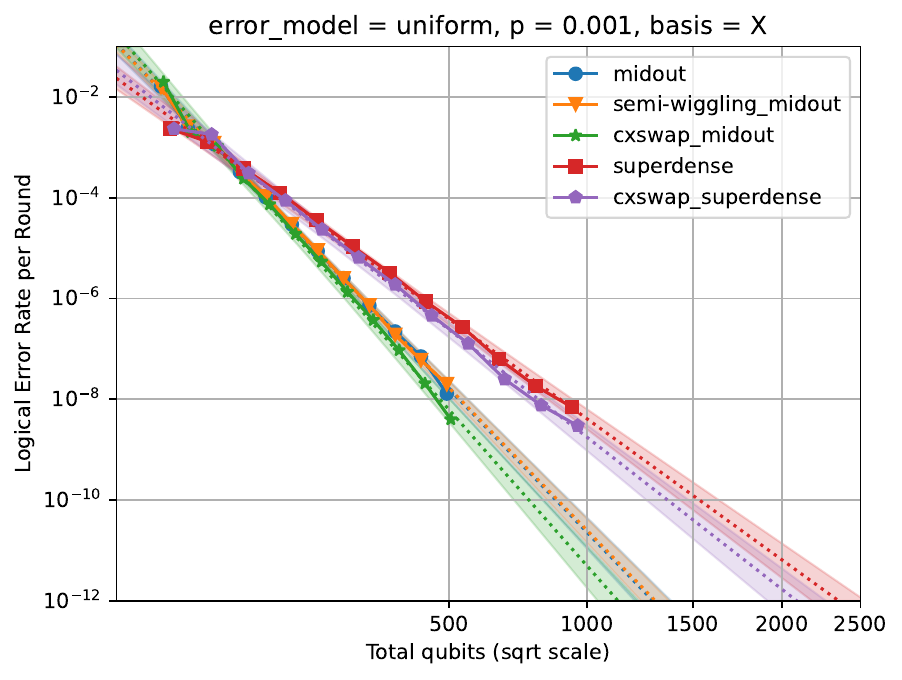}
    \end{minipage}
    \begin{minipage}{0.49\linewidth}
        \includegraphics[width=\linewidth]{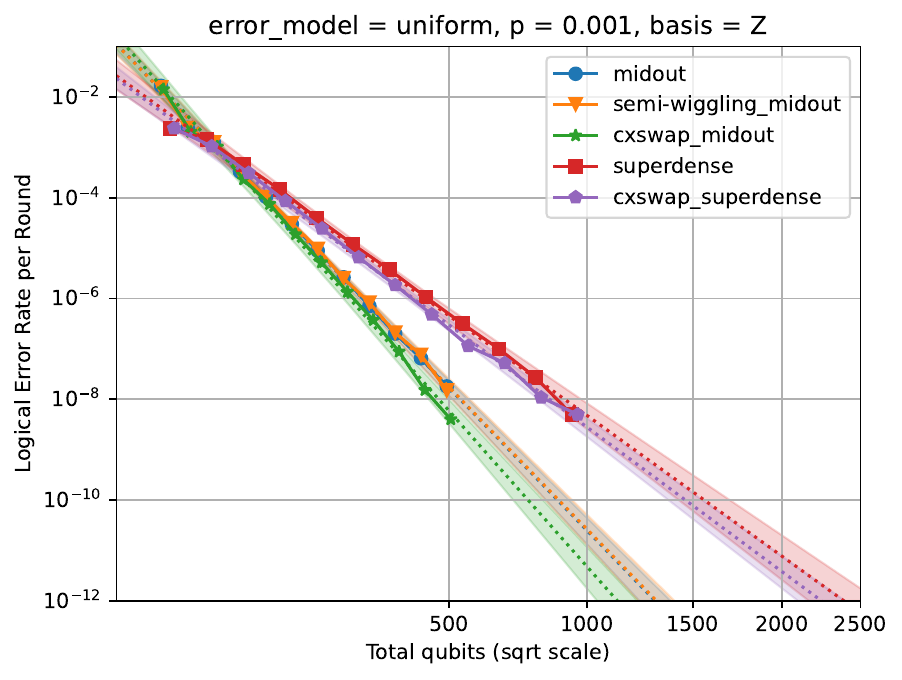}
    \end{minipage}
    \captionsetup{list=no}
    \caption[]{The footprint fitting plot for the physical error rate $p=0.1\%$ using the fitting curve~\eqref{eq:fitting} based on the numerical data shown in Figs.~\ref{fig:logical_error_rate_uniform_X} and \ref{fig:logical_error_rate_uniform_Z}.
    We compare the footprint for the midgut circuit, the semi-wiggling midgut circuit, the CXSWAP midgut circuit, the superdense circuit, and the CXSWAP superdense circuit.
    The fit parameters $a$ and $b$ in Eq.~\eqref{eq:fitting} are estimated by the least square method, and
    the highlighted region corresponds to hypotheses with likelihoods within a factor of 1000 of the maximum likelihood hypothesis, given the sampled data (see Appendix~\ref{appendix_sec:fitting} for the detail).}
    \label{fig:footprint_comparison}
\end{figure}

The threshold plots (logical error rate versus physical error rate) are shown in Figs.~\ref{fig:logical_error_rate_uniform_X} and \ref{fig:logical_error_rate_uniform_Z} in Appendix~\ref{appendix_sec:numerical_result}.
We also estimate the teraquop footprint, which is the required number of qubits to achieve the logical error rate $10^{-12}$.
We plot the logical error rate $P_L$ and the total number of qubits $n$ in Fig.~\ref{fig:footprint_comparison}.
For the color code with distance $d$, the logical error rate scales as
\begin{align}
    P_L\sim p^{\lfloor{d+1\over 2}\rfloor},
\end{align}
and the total number of qubits scales as
\begin{align}
    n= O(d^2).
\end{align}
Therefore, $P_L$ scales as
\begin{align}
    P_L\sim \exp[O(\sqrt{n})].
\end{align}
Based on this observation, we fit the $P_L$ versus $n$ plot by the fitting curve
\begin{align}
\label{eq:fitting}
    \log P_L = a \sqrt{n} + b,
\end{align}
where $a$ and $b$ are fit parameters.
We estimate the fit parameters $a$ and $b$ the least square method to plot the fitting curves shown in Fig.~\ref{fig:footprint_comparison}.
The highlighted region in the plot corresponds to hypotheses with likelihoods within a factor of 1000 of the maximum likelihood hypothesis, given the sampled data (see Appendix~\ref{appendix_sec:fitting} for the detail).

\begin{figure}
    \centering
    \begin{minipage}{0.49\linewidth}
        \includegraphics[width=\linewidth]{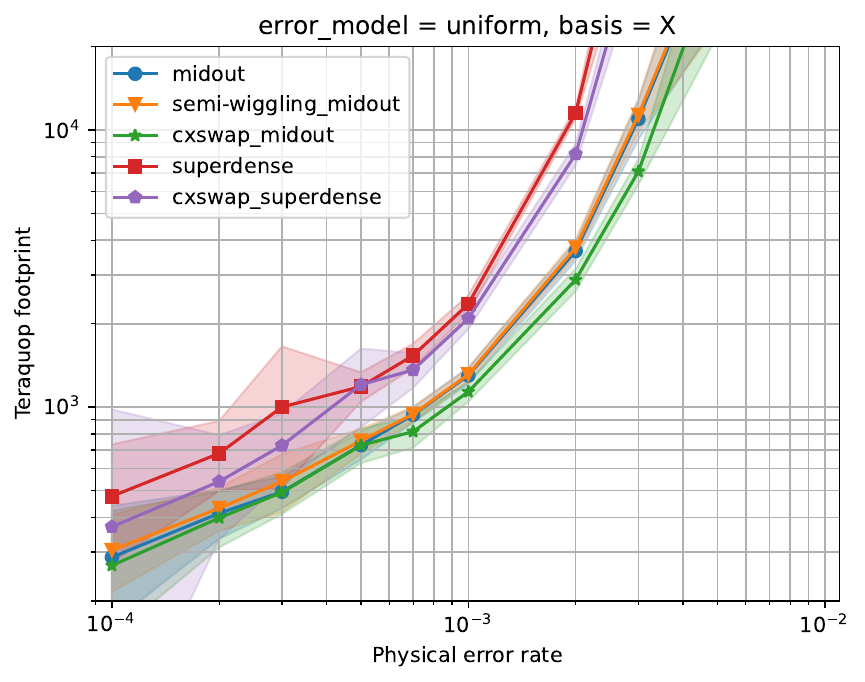}
    \end{minipage}
    \begin{minipage}{0.49\linewidth}
        \includegraphics[width=\linewidth]{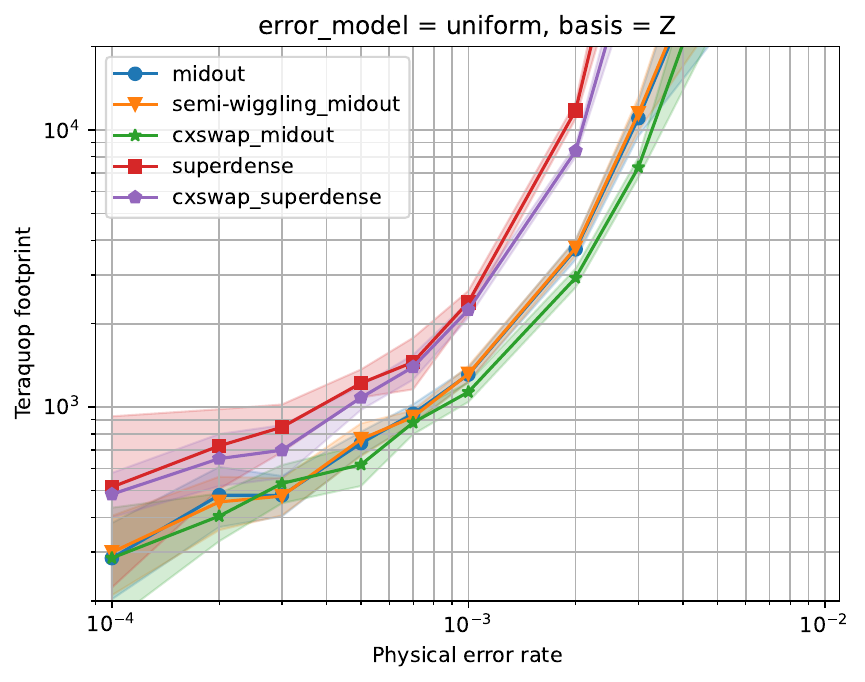}
    \end{minipage}
    \caption{Comparison of the teraquop footprints under the uniform error model with different physical error rates based on the numerical data shown in Figs.~\ref{fig:logical_error_rate_uniform_X} and \ref{fig:logical_error_rate_uniform_Z}.
    The highlighted region corresponds to hypotheses with likelihoods within a factor of 1000 of the maximum likelihood hypothesis, given the sampled data (see Appendix~\ref{appendix_sec:fitting} for the detail).}
    \label{fig:teraquop_footprint_comparison}
\end{figure}

Since the CXSWAP midout (superdense) circuit has a shorter depth than the midout (superdense) circuit, it offers a lower logical error rate, leading to $\sim 10\%$ reduction in the teraquop footprint at the physical error rate $0.1\%$ as shown in Figs.~\ref{fig:footprint_comparison} and \ref{fig:teraquop_footprint_comparison}.
The semi-wiggling midout circuit offers a similar footprint to the original midout circuit since it just changes the direction of some CNOT gates.
However, in experimental settings, it can offer more robustness to leakage error that is not captured in this numerical simulation.

\section{Conclusion}
\label{sec:conclusion}
This work proposes new error correction circuits for the color code: the semi-wiggling midout circuit, the CXSWAP midout circuit, and the CXSWAP superdense circuit.
The semi-wiggling midout circuit enables leakage mitigation schemes that rely on the reset gate by periodically flipping the data and measurement qubit roles of the bulk qubits.
The CXSWAP midout (superdense) circuit has shorter depth than the original midout (superdense) circuit, which results in $\sim 10\%$ reduction in the teraquop footprint under the uniform error model with the physical error rate $p=0.1\%$.
Moreover, the CXSWAP circuits present an additional opportunity to reduce logical error rate for architectures in which two-qubit gates equivalent to CXSWAP can be calibrated to higher flidelity than the more conventional CNOT and CZ gates.

\section{Contributions}
Satoshi built the circuits, simulated their performance, and wrote the paper.
Craig and Matt provided guidance on circuit constructions.
Adam conceived and supervised the project.

\begin{acknowledgments}
This work was supported by Japan Society for the Promotion of Science (JSPS) KAKENHI Grant No. 23KJ0734.
This work was done as an internship project at Google Japan.
\end{acknowledgments}

\appendix

\section{Error model}
\label{appendix_sec:error_model}

In our numerical simulation, we use the uniform error model defined as follows.
The error correction circuits are composed of $\mathrm{Idle}, R_X, R_Z, M_X, M_Z, \mathrm{CNOT}, \mathrm{CXSWAP}$, where $\mathrm{Idle}$ represents the idling gate, $R_X$ ($R_Z$) represents the preparation of $\ket{+}$ ($\ket{0}$) state and $M_X$ ($M_Z$) represents the measurement in the $X$ ($Z$) basis.
These operations are affected by the error channels as shown in Table~\ref{tab:definition_error_model}, where the error channels are defined by
\begin{align}
\label{eq:definition_error_channel}
\begin{split}
    \mathrm{XERR}(p)(\cdot)&\coloneqq (1-p) \cdot + p X\cdot X,\\
    \mathrm{ZERR}(p)(\cdot)&\coloneqq (1-p) \cdot + p Z\cdot Z,\\
    \mathrm{DEP1}(p)(\cdot)&\coloneqq (1-p) \cdot + {p\over 3} X\cdot X + {p\over 3} Y\cdot Y + {p\over 3} Z\cdot Z,\\
    \mathrm{DEP2}(p)(\cdot)&\coloneqq (1-p) \cdot + {p\over 15} \sum_{\substack{P, Q \in \{I,X,Y,Z\} \\ (P,Q) \ne (I,I)}} (P\otimes Q)\cdot (P\otimes Q)
\end{split}
\end{align}
for the physical error rate $p$.

\begin{table}
    \centering
    \begin{tabular}{|c|c|}\hline
        Ideal gate & Noisy gate\\\hline
        Idle & $\mathrm{DEP1}(p)$ \\
        $R_X$ & $\mathrm{ZERR}(p)\circ R_X$\\
        $R_Z$ & $\mathrm{XERR}(p)\circ R_Z$\\
        $M_X$ & $M_X\circ \mathrm{ZERR}(p)$\\
        $M_Z$ & $M_Z\circ \mathrm{XERR}(p)$\\
        $\mathrm{CNOT}$ & $\mathrm{DEP2}(p)\circ \mathrm{CNOT}$\\
        $\mathrm{CXSWAP}$ & $\mathrm{DEP2}(p)\circ \mathrm{CXSWAP}$\\\hline
    \end{tabular}
    \caption{Definition of the uniform error model, where the error channels are defined in Eq.~\eqref{eq:definition_error_channel}.}
    \label{tab:definition_error_model}
\end{table}

\section{Numerical results}
\label{appendix_sec:numerical_result}

\begin{figure}
    \centering
    \includegraphics[width=\linewidth]{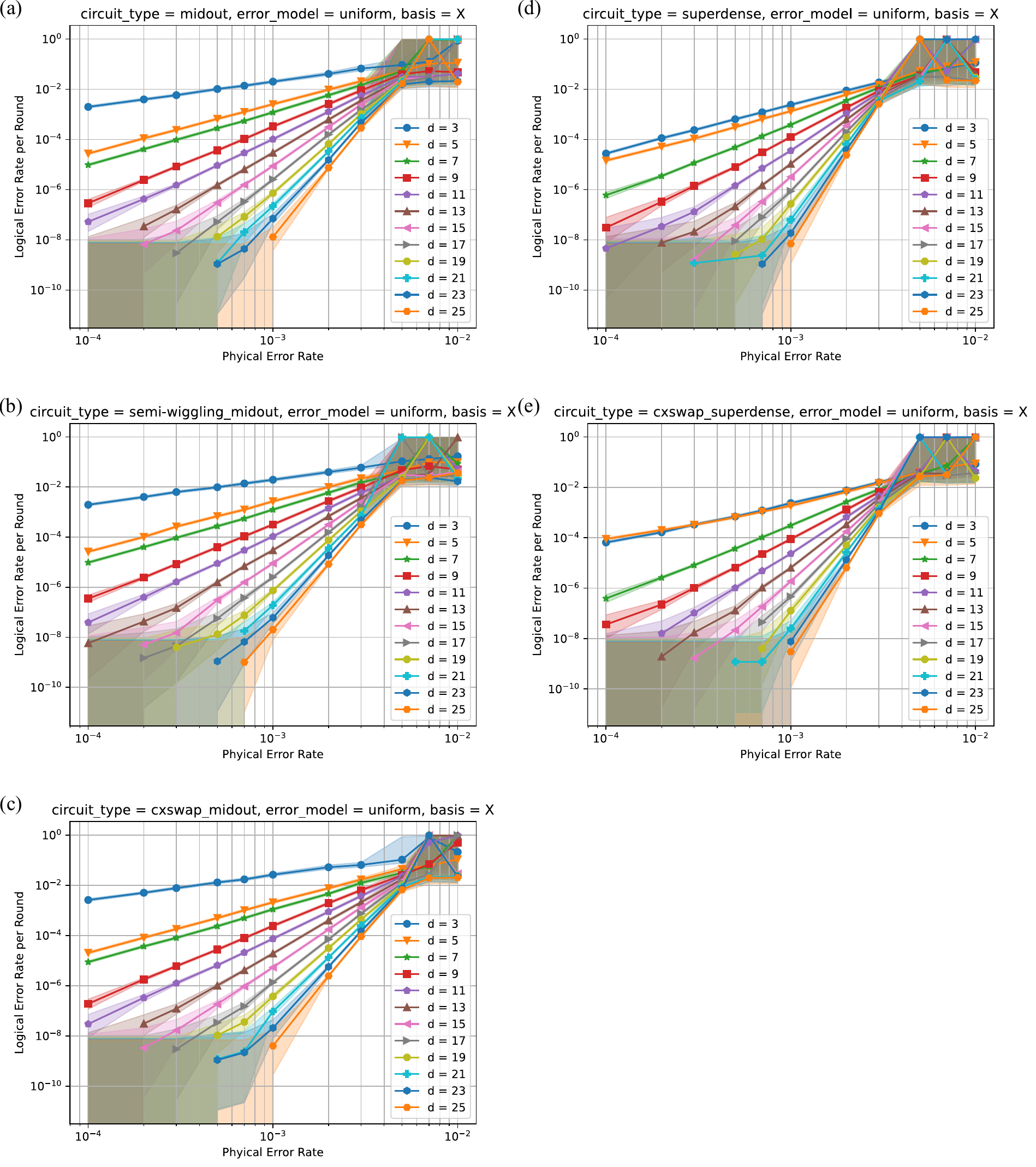}
    \caption{The threshold plots in the $X$-basis simulation for (a) the midout circuit, (b) the semi-wiggling midout circuit, (c) the CXSWAP midout circuit, (d) the superdense circuit, and (e) the CXSWAP superdense circuit.}
    \label{fig:logical_error_rate_uniform_X}
\end{figure}

\begin{figure}
    \centering
    \includegraphics[width=\linewidth]{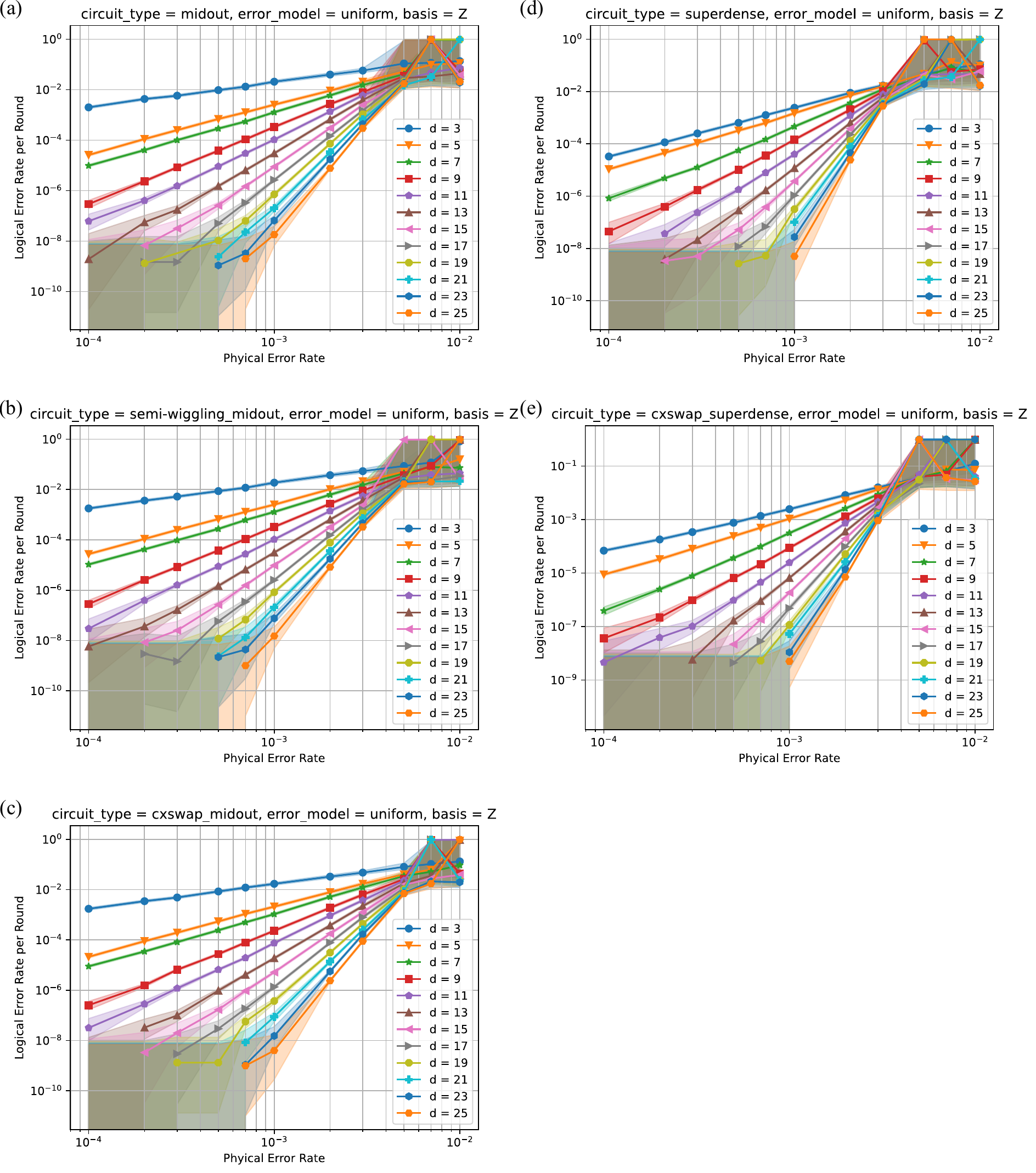}
    \caption{The threshold plots in the $Z$-basis simulation for (a) the midout circuit, (b) the semi-wiggling midout circuit, (c) the CXSWAP midout circuit, (d) the superdense circuit, and (e) the CXSWAP superdense circuit.}
    \label{fig:logical_error_rate_uniform_Z}
\end{figure}

We show the threshold plots in Figs.~\ref{fig:logical_error_rate_uniform_X} and \ref{fig:logical_error_rate_uniform_Z}, whose data are used to plot Figs.~\ref{fig:footprint_comparison} and \ref{fig:teraquop_footprint_comparison} in the main text.

\section{Details of the fitting}
\label{appendix_sec:fitting}
We show the details of the fitting used in Figs.~\ref{fig:footprint_comparison} and \ref{fig:teraquop_footprint_comparison}.
We consider the fitting of $a$ and $b$ in Eq.~\eqref{eq:fitting} based on the sampled data $(n_i, P_{L,i})$ for $i=1,2,\ldots, N$ using the least square method, which minimizes the residual sum of squares defined by
\begin{align}
    J(a,b) \coloneqq \sum_{i=1}^N (\log P_{L,i} - a\sqrt{n_i} - b)^2.
\end{align}
The optimal parameters $a^*, b^*$ are given by those minimizing $J(a,b)$:
\begin{align}
    (a^*, b^*) \coloneqq \arg\min_{a,b} J(a,b).
\end{align}
Assuming that $\log P_{L,i} - a\sqrt{n_i} - b$ follows the normal distribution with mean $0$ and variance $\sigma^2$, the likelihood function $L(a,b)$, defined by the probability to obtain $(a,b)$ given the sampled data $(n_i, P_{L,i})$, is given by
\begin{align}
    L(a,b) \propto \exp\left[-{\sum_{i=1}^{N} \left(\log P_{L,i} - a\sqrt{n_i} - b\right)^2\over 2\sigma^2}\right].
\end{align}
The variance $\sigma^2$ is estimated from the sampled data by
\begin{align}
    \sigma^2 = {1\over N-2} \sum_{i=1}^N (\log P_{L,i} - a^*\sqrt{n_i} - b^*)^2.
\end{align}
Then, the maximum likelihood hypothesis is given by $(a^*, b^*)$.
We consider the set of parameters $(a,b)$ defined by
\begin{align}
    \left\{(a,b) \; \middle| \; L(a,b) \geq {1\over 1000} L(a^*, b^*)\right\}
    = \left\{(a,b) \; \middle| \; J(a,b) \leq J(a^*, b^*)+ 2\ln(1000) \sigma^2\right\},
\end{align}
which corresponds to the highlighted region in Figs.~\ref{fig:footprint_comparison} and \ref{fig:teraquop_footprint_comparison}.

\bibliographystyle{apsrev4-2}
\bibliography{main}

\end{document}